\UseRawInputEncoding
\documentclass[aip,jap,reprint]{revtex4-1}

\usepackage{graphicx}
\usepackage{dcolumn}
\usepackage{bm}

\usepackage{amsmath}
\usepackage[utf8]{inputenc}
\usepackage[T1]{fontenc}
\usepackage{mathptmx}
\usepackage{etoolbox}
\usepackage{siunitx}
\usepackage{amssymb}
\usepackage{color,soul}
\usepackage{svg}
\usepackage{natbib}
\usepackage{microtype}
\usepackage{textcomp}
\usepackage{array}
\usepackage{tabularx}
\usepackage{tikz}
\usepackage{multirow}

\definecolor{blue0}{HTML}{3853A4}
\definecolor{green0}{HTML}{0D8040}
\definecolor{red0}{HTML}{ED2224}

\makeatletter
\def\@email#1#2{%
 \endgroup
 \patchcmd{\titleblock@produce}
  {\frontmatter@RRAPformat}
  {\frontmatter@RRAPformat{\produce@RRAP{*#1\href{mailto:#2}{#2}}}\frontmatter@RRAPformat}
  {}{}
}%
\makeatother
\begin{document}

\preprint{AIP/123-QED}

\title{
Two-Carrier Model-Fitting of Hall Effect in Semiconductors with Dual-Band Occupation: A Case Study in GaN Two-Dimensional Hole Gas
}
\author{Joseph E. Dill*}
 \affiliation{School of Applied \& Engineering Physics, Cornell University, Ithaca, New York 14853, USA}
\author{Chuan F. C. Chang*}%
\affiliation{Department of Physics, Cornell University, Ithaca, New York 14853, USA}%
\author{Debdeep Jena}
\author{Huili Grace Xing}
\affiliation{Electrical and Computer Engineering, Cornell University, Ithaca, New York 14853, USA}
\affiliation{Materials Science and Engineering, Cornell University, Ithaca, New York 14853, USA}
\affiliation{Kavli Institute at Cornell for Nanoscale Science, Cornell University, Ithaca, New York 14853, USA\\ *These two authors contributed equally to this work.}
\date{\today}

\begin{abstract}
We develop a two-carrier Hall effect model fitting algorithm to analyze temperature-dependent magnetotransport measurements of a high-density ($\sim\SI{4e13}{\per\centi\meter\squared}$) polarization-induced two-dimensional hole gas (2DHG) in a GaN/AlN heterostructure.
Previous transport studies in GaN 2DHGs have reported a two-fold reduction in 2DHG carrier density from room to cryogenic temperature. We demonstrate that this apparent drop in carrier density is an artifact of assuming one species of carriers when interpreting Hall effect measurements.
Using an appropriate two-carrier model, we resolve light hole (LH) and heavy hole (HH) carrier densities congruent with self-consistent Poisson-k$\cdot$p simulations and observe an LH mobility of $\sim$1400 cm$^2$/Vs and HH mobility of $\sim$300 cm$^2$/Vs at 2 K. 
This report constitutes the first experimental signature of LH band conductivity reported in GaN.

\end{abstract}

\maketitle

\begin{figure}
    \centering
    \includegraphics[]{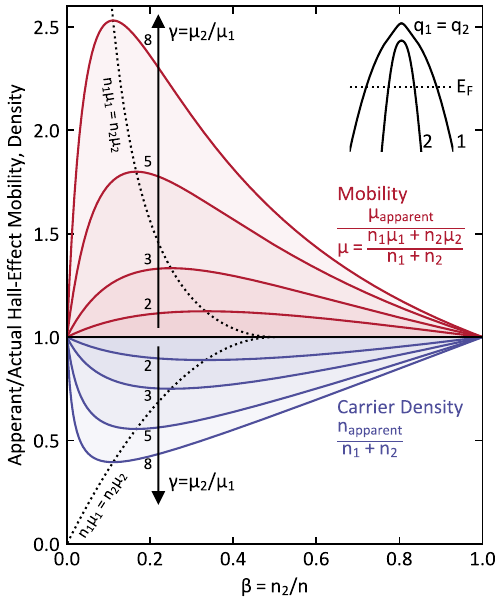}
    \caption{Ratio between the apparent Hall mobility and true mobility $\mu_{apparent}/\mu$ (red), and likewise for the carrier density $n_{apparent}/n$ (blue), obtained from a single-carrier interpretation of low-field Hall effect measurements of a system of two conducting channels with the same charge polarity ($q_1=q_2$): $n=n_1+n_2$; $\mu=(n_1\mu_1+n_2\mu_2)/n$. Both are plotted as a function of the fraction of carrier in the 2nd channel $\beta=n_2/n$ and the ratio of mobilities between the two channels $\gamma=\mu_2/\mu_1$. Errors peak when the conductivities of both channels are equal ($n_1\mu_1=n_2\mu_2$).}
    \label{fig: Hall_error}
\end{figure}

\section{Introduction}

P-type polarization doping in metal-polar GaN/Al(Ga)N heterostructures has been developed \cite{Shur2000} as an alternative to chemical doping, which suffers from the lack of a shallow acceptor dopant in GaN (activation energy 135$\sim$170 meV for Mg\cite{Kim1996, Gotz1996, Kim2000, Chaudhuri2019}). Polarization discontinuity at the hetero-interface, combined with piezoelectric charge from epitaxial strain of the GaN layer, can produce a two-dimensional hole gas (2DHG) confined to the GaN\cite{Ambacher1999}, even without intentional chemical dopants, as recently demonstrated by \citet{Chaudhuri2019} and \citet{Beckmann2022}.

In a singular undoped GaN/AlN heterojunction\cite{Chaudhuri2019, Chaudhuri2021}, the 2DHG density is $\sim\SI{4e13}{\per\centi\meter\squared}$. The lack of parasitic electron channels and ionized impurities from chemical dopants in this material stack greatly minimizes extrinsic carrier scattering, boosting cryogenic hole mobility. Recent growths of GaN/AlN 2DHGs by molecular beam epitaxy (MBE) have also utilized buried impurity blocking layers\cite{Chaudhuri2021} and single-crystal AlN substrates\cite{Zhang2021} to reduce impurity and dislocation scattering, respectively, enabling a sheet resistance in a GaN 2DHG below 1 k$\Omega$/$\square$, and reported mobility above 280 cm$^2$/Vs at 10 K\cite{Zhang2021}. This high mobility (relative to what has previously been reported in p-GaN) makes this material system well-suited to study hole transport phenomenon in GaN. However, the high carrier density in the GaN/AlN 2DHG presents experimental complications in accurately characterizing its transport.

Classical Hall effect measurements\cite{Hall1879} are a standard means of characterizing transport in semiconductors. With a source current $I$, the sheet resistance is determined from a zero magnetic field measurement of the longitudinal voltage $V_x$, 
\begin{equation}\label{eq: Rsh}
    R_{sh} = fR_{xx} = f\frac{V_x}{I} = \frac{1}{qn\mu},
\end{equation}
($q$ is the electron charge; $f$ is a geometric factor) and the transverse voltage $V_y$ under a perpendicular magnetic field $B$ gives the Hall coefficient,
\begin{equation}\label{eq: RH}
    R_{H} = \frac{R_{xy}}{B} = \frac{V_y}{IB} = \frac{1}{qn}.
\end{equation}
These equations decouple the free carrier density $n=1/qR_H$ and mobility $\mu=R_H/R_{sh}$ in terms of the measured quantities. Because $V_{y}$ is expected to vary linearly with $B$ (see Eq. \ref{eq: RH}), most quick-feedback Hall effect measurement systems only measure $V_y$ at one or two low magnetic field points to calculate $R_{xy}$ (typically $\sim\SI{0.5}{T}$ for a tabletop Hall setup). 

However, these equations assume a single carrier population with uniform mobility. Due to its high carrier density, self-consistent Poisson-k$\cdot$p simulations of the GaN/AlN 2DHG place the Fermi level 40$\sim$60 meV below the heavy hole (HH) $\Gamma$-point, and 25$\sim$35 meV below the light hole (LH) $\Gamma$-point, degenerately occupying both bands\cite{Bader2019Wurtzite} (see bandstructure inset in Fig. \ref{fig: Hall_error}). The LH band also exhibits a higher group velocity than the HH band, so the LH and HH carriers conduct in parallel, with different densities and mobilities. In such cases, $R_{xy}(B)$ and $R_{xx}(B)$ become nonlinear in $B$ and must be analyzed with a Hall effect model that accounts for the plurality of carrier species.  

This report applies a classical Drude model for isotropic magnetotransport of a two-carrier system. It then outlines a fitting procedure by which the total carrier density and mobility can be estimated from Hall effect measurements taken with sufficiently high magnetic field to capture nonlinear field dependence in $R_{xx}(B)$ and $R_{xy}(B)$. In the case where the two carriers populations have the same charge polarity ($q_1=q_2$), we demonstrate that a single-carrier interpretation (Eqs. \ref{eq: Rsh} and \ref{eq: RH}) of low magnetic field measurements will always underestimate the total carrier concentration (thus overestimating the mobility). This is shown in Fig. \ref{fig: Hall_error} and discussed in Sec. \ref{subsec: error in single-carrier}.

This insight elucidates previous reports of temperature-dependent Hall effect measurements in (In)GaN/Al(Ga)N 2DHGs\cite{Liu2013, Li2013, Chaudhuri2019, Zhang2021, Chaudhuri2022a, Chaudhuri2022, Beckmann2022}, all of which utilize a single-carrier interpretation for low-field measurements, and report a roughly two-fold decrease in hole density between room and cryogenic temperature. In these reports, this temperature dependence is either ignored or temporarily attributed to dopant activation or temperature dependence of the GaN-air surface barrier height, piezoelectric constants, or spontaneous polarizations of the constituent materials. However, no model invoking these mechanisms has been adequate to account for this magnitude of change in carrier density. Using the two-carrier model, we demonstrate that this apparent reduction in carrier density can be explained by an increase in the LH-to-HH mobility ratio with decreasing temperature. 

\section{The Two-Carrier Transport Model}\label{sec: The Two-Carrier Transport Model}

The isotropic conductivity $\sigma$ of $m$ parallel-conducting carrier populations subject to a perpendicular magnetic field $B$ can be expressed as \cite{Petritz1958}
\begin{subequations}\label{eq: sigma}
\begin{align}
    \sigma_{xx}(B) &= \sum_{i=1}^m \frac{\sigma_i}{1+(\mu_i B)^2} = \sum_{i=1}^m \frac{q_in_i\mu_i}{1+(\mu_i B)^2},\label{eq: sigma_xx} \\
    \sigma_{xy}(B) &= \sum_{i=1}^m \frac{\sigma_i(\mu_i B)}{1+(\mu_i B)^2} = \sum_{i=1}^m \frac{q_in_i\mu_i^2B}{1+(\mu_i B)^2},\label{eq: sigma_xy}
\end{align}
\end{subequations}
with the convention that $q_i$ and $\mu_i$ carry the same sign \cite{Kim1993}. From the conductivity tensor equation, 
\begin{equation}\label{eq: sigma tensor}
    \mathbf{J} = \begin{pmatrix}
    \sigma_{xx} & \sigma_{xy}\\
    -\sigma_{xy} & \sigma_{xx}
    \end{pmatrix} \mathbf{E} \Longleftrightarrow \begin{pmatrix}
    R_{xx} & R_{xy}\\
    -R_{xy} & R_{xx}
    \end{pmatrix} \mathbf{J} = \mathbf{E},
\end{equation}
the longitudinal and transverse resistances are expressed as
\begin{subequations}\label{eq: Rxx,xy}
\begin{align}
    R_{xx}(B) &= \frac{\sigma_{xx}(B)}{\sigma_{xx}^2(B) + \sigma_{xy}^2(B)}, \label{eq: Rxx}\\
    R_{xy}(B) &= \frac{\sigma_{xy}(B)}{\sigma_{xx}^2(B) + \sigma_{xy}^2(B)}\label{eq: Rxy}.
\end{align}
\end{subequations}
In the case of a single carrier species ($m=1$), Eqs. \ref{eq: Rsh} and \ref{eq: RH} are recovered. In the case of two carrier species, the resistance tensor components become
\begin{widetext}
\begin{subequations}\label{eq: Rxx,xy_nmu}
\begin{align}
R_{xx} &= \frac{\sigma_1\big(1+(\mu_2B)^2\big) + \sigma_2\big(1+(\mu_1B)^2\big)}{\sigma_1^2\big(1+(\mu_2B)^2\big) + 2\sigma_1\sigma_2\big(1+(\mu_1\mu_2B^2)\big) + \sigma_1^2\big(1+(\mu_2B)^2\big)};\label{eq: Rxx_nmu}\\
R_{xy} &= \frac{\sigma_1(\mu_1B)\big(1+(\mu_2B)^2\big) + \sigma_2(\mu_2B)\big(1+(\mu_1B)^2\big)}{\sigma_1^2\big(1+(\mu_2B)^2\big) + 2\sigma_1\sigma_2\big(1+(\mu_1\mu_2B^2)\big) + \sigma_1^2\big(1+(\mu_2B)^2\big)}.\label{eq: Rxy_nmu}
\end{align}
\end{subequations}
\end{widetext}
This tensor-style derivation of these equations is given in many places \cite{Ashcroft1976, Werner2017, Kim1993, Petritz1958}. Example use cases include parallel conduction in bulk and surface states \cite{Petritz1958, Czaya1977}, parallel conduction of electrons and holes \cite{Soule1958}, multi-sub-band occupation in quantum well heterostructures \cite{Zaremba1992}, and dual-band occupation in high-density n-type \cite{Khan1992} and p-type \cite{Kim1996b} semiconductors. In any case, the mobility and density of each conducting channel are determined as free fitting parameters.

More complex Quantitative Mobility Spectrum Analysis (QMSA)  algorithms have been developed \cite{Beck1987, Beck2021} which generalize Eqs. \ref{eq: sigma} to account for non-uniform group velocity and carrier scattering within occupied bands. This is done by replacing the discrete densities $n_i$ with "mobility spectra" $n(\mu)$ and making $m$ large such that the summations approximate integration, significantly increasing the number of free fitting parameters. Such models have been applied previously to characterize magnetotransport in various n-type III-Nitride semiconductor systems \cite{Biyikli2006, Fehlberg2006, Lisesivdin2007}. 

However, the simpler four-parameter two-carrier model employed in this paper proves adequate for distinguishing between HH and LH band transport. Further, this model is also shown to overfit measurements above $\sim$100 K due to the lack of polynomial features in the fitted data; thus, employing a model with even more free parameters is unwarranted.

\subsection{Polynomial expansion coefficients of $R_{xx}(B)$ and $R_{xy}(B)$}\label{subsec: Polynomial expansion}
\begin{figure*}
    \centering
    \includegraphics[width=\textwidth]{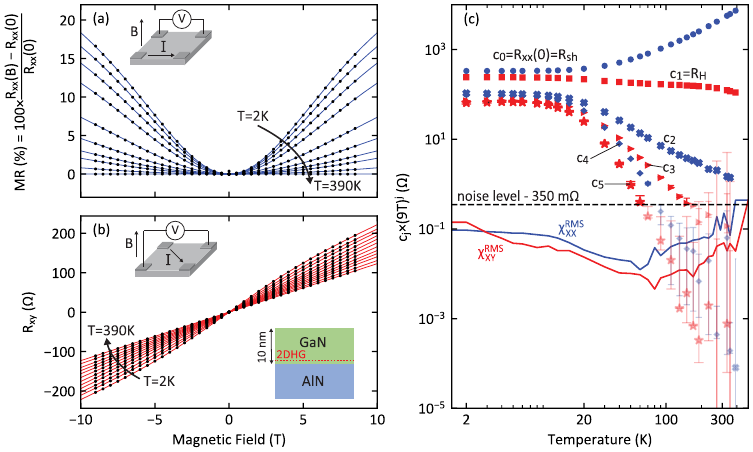}
    \caption{Measured (a) MR$(B)$ and (b) $R_{xy}(B)$ of a GaN/AlN 2DHG, field-swept to $B=\pm$\SI{9}{T} at select temperatures from \SIrange{2}{390}{K}. The four-point measurement configuration is shown for each. (c) The RMS residuals of the fits in (a) $\chi_{xx}^{RMS}$ (blue line) and (b) $\chi_{xy}^{RMS}$ (red line) versus temperature, and the signal contributions of the $j$-th polynomial features of $R_{xx}(B)$ (even $n$) or $R_{xy}(B)$ (odd $n$) at $B$=9 T (the maximum measured field) for $n=0$ (circles), $n=1$ (squares), $n=2$ (X's), $n=3$ (triangles), $n=4$ (diamonds), and $n=5$ (stars). Points with an uncertainty bound below the \SI{350}{\milli\ohm} noise line are dimmed.
    }
    \label{fig: mr_data_cj}
\end{figure*}

Valuable insight and restructuring of Eqs. \ref{eq: Rxx,xy_nmu} is obtained by expanding in powers of $B$,
\begin{widetext}
\begin{subequations}\label{eq: Rxx,xy_expand}
\begin{align}
    R_{xx}(B) &\approx \overbrace{\frac{1}{q_1n_1\mu_1+q_2n_2\mu_2}}^{c_0=\frac{1}{\sigma}=R_{sh}} + \overbrace{\frac{q_1q_2n_1n_2\mu_1\mu_2(\mu_1-\mu_2)^2}{(q_1n_1\mu_1+q_2n_2\mu_2)^3}}^{c_2}B^2  - \overbrace{\frac{q_1q_2n_1n_2\mu_1^3\mu_2^3(q_1n_1+q_2n_2)^2(\mu_1-\mu_2)^2}{(q_1n_1\mu_1+q_2n_2\mu_2)^5}}^{c_4=c_3^2/c_2}B^4 + \dots, \label{eq: Rxx_expanded}\\
    R_{xy}(B) &\approx \underbrace{\frac{q_1n_1\mu_1^2+q_2n_2\mu_2^2}{(q_1n_1\mu_1+q_2n_2\mu_2)^2}}_{c_1=R_H}B - \underbrace{\frac{q_1q_2n_1n_2\mu_1^2\mu_2^2(\mu_1-\mu_2)^2}{(q_1n_1\mu_1+q_2n_2\mu_2)^4}}_{c_3}B^3 + \underbrace{\frac{q_1q_2n_1n_2\mu_1^4\mu_2^4(q_1n_1+q_2n_2)^3(\mu_1-\mu_2)^2}{(q_1n_1\mu_1+q_2n_2\mu_2)^6}}_{c_5=c_3^3/c_2^2}B^5 - \dots \label{eq: Rxy_expanded}.
\end{align}
\end{subequations}
\end{widetext}
Here, $R_{xx}(B)$ and $R_{xy}(B)$ are seen to have strictly even and odd polynomial magnetic field dependence, respectively, with polynomial coefficients that alternate signs.

The $c_j$ expansion coefficients in Eqs. \ref{eq: Rxx,xy_expand} satisfy a recursive condition, 
\begin{equation}\label{eq: cj}
    c_{j\geq2} = \frac{c_3^{j-2}}{c_2^{j-3}} = \frac{R_H ^j}{R_{sh} ^{j-1}}\underbrace{\frac{\beta(1-\beta)(\gamma-1)^2\gamma^{j-1}}{\big(1+\beta(\gamma^2+1)\big)^j}}_{\sim1},
\end{equation}
such that $R_{xx}(B)$ and $R_{xy}(B)$ in Eqs. \ref{eq: Rxx,xy_nmu} are equivalently expressed (without truncating the expansions) as 
\begin{subequations}\label{eq: Rxx,xy_cj}
\begin{align}
    R_{xx}(B) &= c_0 + \frac{c_2B^2}{1+\big(\frac{c_3}{c_2}B\big)^2} \label{eq: Rxx_cj} \\
    R_{xy}(B) &= c_1B - \frac{c_3B^3}{1+\big(\frac{c_3}{c_2}B\big)^2}.\label{eq: Rxy_cj}
\end{align}
\end{subequations}

$c_{0-3}$ are better-suited as fitting parameters for $R_{xx}(B)$ and $R_{xy}(B)$ measurements (than, for example, $\sigma_1$, $\sigma_2$, $n_1$, and $n_2$ in Eqs. \ref{eq: Rxx,xy}) for three reasons: 1) Each fitting parameter pertains to a unique polynomial feature of the data, minimizing covariance with $c_0$ or $c_1$; 2) The covariance between $c_2$ and $c_3$ only influences how well the fit captures the $B^{>4}$ features of the $R_{xx}(B)$ and $R_{xy}(B)$ data (see Eq. \ref{eq: Rxx,xy_expand}); 3) This parameterization guards against over-fitting. If $n_1$, $n_2$, $\mu_1$, and $\mu_2$ are not known \textit{apriori}, then four free fitting parameters are required, and $R_{xx}(B)$ and $R_{xy}(B)$ must collectively exhibit four or more polynomial features to constrain these parameters uniquely. If the $c_3B^3$ feature in $R_{xy}(B)$ is poorly resolved above the measurement noise, $c_3$ will be unconstrained, signaling over-fitting.

The band-resolved mobilities and densities are obtained from the $c_{0-3}$ coefficients by
\begin{subequations}\label{eq: nmu from cj}
\begin{align}
    n_1 &= \frac{q_1}{q_2} \frac{2c_2^3 - c_1c_2c_3 + c_3(c_0c_3 + \text{sgn}(q_2)c^*)}{2q(c_1c_3-c_2^2)c^*} \label{eq: n1 from cj}\\
    n_2 &= - \frac{2c_2^3 - c_1c_2c_3 + c_3(c_0c_3 - \text{sgn}(q_2)c^*)}{2q(c_1c_3-c_2^2)c^*} \label{eq: n2 from cj}\\
    \mu_{_2^1} &= \frac{c_1c_2+c_0c_3\mp \text{sgn}(q_2) c^*}{2c_0c_2} \label{eq: mu from cj}
\end{align}
\end{subequations}
where
\begin{equation}
    c^* = \sqrt{c_2^2(c_1^2+4c_0c_2)-2c_0c_1c_2c_3+c_0^2c_3^2}.
\end{equation}
These equations hold for any combination of signs in $q_1$ and $q_2$.

In Sec. \ref{subsec: Fitting with cj}, this polynomial coefficient representation of the two-carrier model is applied to temperature-dependent magnetotransport measurements of a GaN/AlN 2DHG.

\subsection{Error in single-carrier analysis of low-field Hall (for $q_1=q_2$)}\label{subsec: error in single-carrier}

We now assess the nature of errors that arise from single-carrier Hall analysis of a system of two carriers with the same charge polarity ($q_1=q_2$), as is the case for the GaN/AlN 2DHG discussed in Sec. \ref{sec: Measurements and analysis}.

From the first-order term of $R_{xy}(B)$ in Eq. \ref{eq: Rxy_expanded} (with $q_1=q_2$), we find that low-field Hall effect measurements (where $B^{2+}$ terms are imperceptible) that are interpreted with a single-carrier model (see Eq. \ref{eq: RH}) will infer an \textit{apparent} carrier density of
\begin{equation}\label{eq: n_ap}
    n_{apparent} = \frac{1}{qR_H} = n\frac{(1+\beta(\gamma-1))^2}{1+\beta(\gamma^2-1)},
\end{equation}
where $n = n_1+n_2$ is the true carrier density, $\beta=n_2/n$ is the fraction of carriers in the 2nd channel, and $\gamma=\mu_2/\mu_1$ is the ratio between the two mobilities. The apparent mobility similarly differs according to 
\begin{equation}\label{eq: mu_ap}
     \mu_{apparent}=\frac{R_H}{R_{sh}} = \mu\frac{1+\beta(\gamma^2-1)}{(1+\beta(\gamma-1))^2},
\end{equation}
where $\mu$ is the ensemble mobility,
\begin{equation}
    \mu = \frac{\sigma}{qn} = \frac{\sigma_1+\sigma_2}{q(n_1+n_2)} = \frac{n_1\mu_1+n_2\mu_2}{n_1+n_2}.
\end{equation}
The ratios $n_{apparent}/n$ and $\mu_{apparent}/\mu$ are plotted in Fig. \ref{fig: Hall_error}, as functions of $\beta$ and $\gamma$. This plot demonstrates that single-carrier interpretations will always \textit{under}-estimate total free carrier density and, consequently, \textit{over}-estimate ensemble mobility. These errors magnify with increasing $\mu_2/\mu_1$ and are highest when $\sigma_1=\sigma_2$. As expected, the measurement errors vanish for single-carrier-equivalent conditions: $\mu_2/\mu_1=1$, $n_2/n=0$, or $n_2/n=1$. 

While this model is applied in this report to a degenerate hole gas (example bandstructure plotted in the upper right corner of Fig. \ref{fig: Hall_error}), the plotted errors apply to any system with isotropic parallel transport of two species with the same charge polarity.

\section{Measurements and analysis}\label{sec: Measurements and analysis}

The sample investigated in this report consisted of a \SI{10}{\nano\meter} thin film of undoped GaN on \SI{700}{\nano\meter} of undoped AlN, grown by molecular beam epitaxy on a single-crystal aluminum nitride substrate (see Fig. \ref{fig: mr_data_cj}c). The details of the growth process are discussed elsewhere\cite{Zhang2021, Chaudhuri2022a}.

Magnetotransport measurements were performed in a Quantum Design PPMS\textsuperscript{\textregistered} DynaCool\textsuperscript{\texttrademark} on a square as-grown sample with soldered indium contacts in van der Pauw geometry. An excitation current of \SI{100}{\micro\ampere} was used at temperatures from \SIrange{2}{390}{\kelvin}, sweeping magnetic field to $\pm$\SI{9}{\tesla}. $R_{xy}$ was measured along a single orientation, while $R_{xx}$ was computed by the van der Pauw equation from two orthogonal measurements\cite{Werner2017}. $R_{xx}(B)$ measurements were symmetrized in $B$, and $R_{xy}(B)$ was anti-symmetrized to account for thermal gradients and contact misalignment\cite{Look1989}. The measured magnetoresistance, $MR(B)=\frac{R_{xx}(B)-R_{xx}(0)}{R_{xx}(0)}$ and $R_{xy}(B)$, are plotted in Figs. \ref{fig: mr_data_cj}a and \ref{fig: mr_data_cj}b, respectively.

\subsection{Fitting $R_{xx}(B)$ and $R_{xy}(B)$ with the $c_j$ coefficients}\label{subsec: Fitting with cj}
\begin{table*}
\caption{\label{tab: fit}Specifications for the three least-squares fits performed in this study.}
\begin{ruledtabular}
\begin{tabular}{c c c c c r@{\extracolsep{0cm}}l}
\multirow{2}{*}{Fit Name} & \multirow{2}{*}{Color} & Fitted & Free & Constrained &  \multicolumn{2}{c}{\multirow{2}{*}{Minimized Quantity}} \\
 &  & Data & Parameters & Parameters &  \\
\hline
XX & \tikz\draw[blue0,fill=blue0] (0,0) circle (.5ex); & $(B^i, R_{xx}^i)$ & $c_0$, $c_2$, $c_3$ & $c_1$ - from XY & $\displaystyle\chi_{xx}^{RMS}$ &$\displaystyle= \sqrt{\frac{\sum_i^N\chi_{xx}^i(c_0,c_2,c_3)^2}{N-3}}$ \\
XY & \tikz\draw[red0,fill=red0] (0,0) circle (.5ex); & $(B^i, R_{xy}^i)$ & $c_1$, $c_2$, $c_3$ & $c_0$ - from XX & $\displaystyle\chi_{xy}^{RMS}$ &$\displaystyle= \sqrt{\frac{\sum_i^N\chi_{xy}^i(c_1,c_2,c_3)^2}{N-3}}$ \\
Simultaneous & \tikz\draw[green0,fill=green0] (0,0) circle (.5ex); & $(B^i, R_{xx}^i, R_{xy}^i)$ & $c_0$, $c_1$, $c_2$, $c_3$ & - & $\displaystyle\chi_{sim}^{RMS}$ &$\displaystyle= \sqrt{\frac{\sum_i^N\chi_{xx}^i(c_0,c_2,c_3)^2+\chi_{xy}^i(c_1,c_2,c_3)^2}{2N-4}}$ \\
\end{tabular}
\end{ruledtabular}
\end{table*}

Measured $R_{xx}(B)$ and $R_{xy}(B)$ were fit to Eqs. \ref{eq: Rxx_cj} and \ref{eq: Rxy_cj}, respectively, with $c_{0-3}$ as free fitting parameters. Three different least-squares fits were performed (summarized in Table \ref{tab: fit}):
\begin{itemize}
    \item Measured $R_{xx}(B)$ were fit independently with $c_0$, $c_2$, and $c_3$ as free parameters (the "XX fit")
    \item Measured $R_{xy}(B)$ were fit independently with $c_1$, $c_2$, and $c_3$ as free parameters (the "XY fit")
    \item $R_{xx}(B)$ and  $R_{xy}(B)$ were fit simultaneously with $c_0$, $c_1$, $c_2$, $c_3$ as free parameters (the "simultaneous fit")
\end{itemize}
The residuals between the model and the measured data points $(B^i, R_{xx}^i, R_{xy}^i)$ were computed by
\begin{subequations}
\begin{align}
    \chi_{xx}^i(c_0,c_2,c_3) &= R_{xx}^i-R_{xx}(B_i;c_0,c_2,c_3), \\
    \chi_{xy}^i(c_1,c_2,c_3) &= R_{xy}^i-R_{xy}(B_i;c_1,c_2,c_3).
\end{align}
\end{subequations}
For each fit, the free parameters were varied to minimize the RMS residual $\chi^{RMS}$ (equations given in Table \ref{tab: fit}). While all three fits should, in theory, yield equivalent band-resolved HH mobilities and densities, fitting $R_{xx}(B)$ and $R_{xy}(B)$ in isolation enables us to assess the degree to which the nonlinearity in each is described by parallel conduction vs some other source of magnetoresistance.

The model results of the XX and XY fits are plotted against the measured data in Figs. \ref{fig: mr_data_cj}a and \ref{fig: mr_data_cj}b, respectively. The corresponding values of $\chi_{xx}^{RMS}$ and $\chi_{xy}^{RMS}$ at each temperature are plotted in Fig. \ref{fig: mr_data_cj}c, ranging from $\sim$\SIrange{10}{100}{\ohm}. Also plotted at each temperature is $c_j \times (\SI{9}{\tesla})^j$, the $j^{\text{th}}$-order contribution to $R_{xx}$ (even $j$) or $R_{xy}$ (odd $j$) at $B=\SI{9}{\tesla}$. $c_0$ (or $R_{sh}$) increases from 0.33 k$\Omega$/$\square$ at \SI{2}{\kelvin} to 5.1 k$\Omega$/$\square$ at room temperature. $c_1$ (or $R_H$) decreases slightly from 1/(\num{2.3e13}) cm$^3$/q at \SI{2}{\kelvin} to 1/(\num{4.4e13}) cm$^3$/q at room temperature, consistent with the apparent decrease in carrier density seen in previous reports of GaN/AlN 2DHG transport \cite{Liu2013, Chaudhuri2019, Zhang2021, Chaudhuri2021}.
For all $j>2$, the magnitude of the $j^{\text{th}}$-order contribution is suppressed monotonically with increasing temperature in conjunction with the increase in $R_{sh}$ (see Eq. \ref{eq: cj}).

Below \SI{350}{\milli\ohm}, $c_3$ ceases changing monotonically with temperature, as anticipated from Eq. \ref{eq: cj} and the observed trend in $c_0$. Fits at each temperature are only considered valid if the lower uncertainty bound of all four free parameters lies above
this noise line, as any nonlinearity observed in $R_{xy}(B)$ is not otherwise convincingly attributable to two-carrier phenomena. The three fits are thus cut off between 70 K and 170 K.

The error bars in Fig. \ref{fig: mr_data_cj}c indicate the range of parameters for which $\chi^{RMS}$ differs from the minimum by $\leq10\%$. This uncertainty range is for larger $j$ and increases at high temperatures where these polynomial features lose prominence in the measured data. The error bounds are discussed further in Sec. \ref{subsec: fitting and error}.

\subsection{The HH and LH Densities \& Mobilities vs Temperature}\label{subsec: HH and LH mobility}
\begin{figure*}
    \centering
    \includegraphics[width=\textwidth]{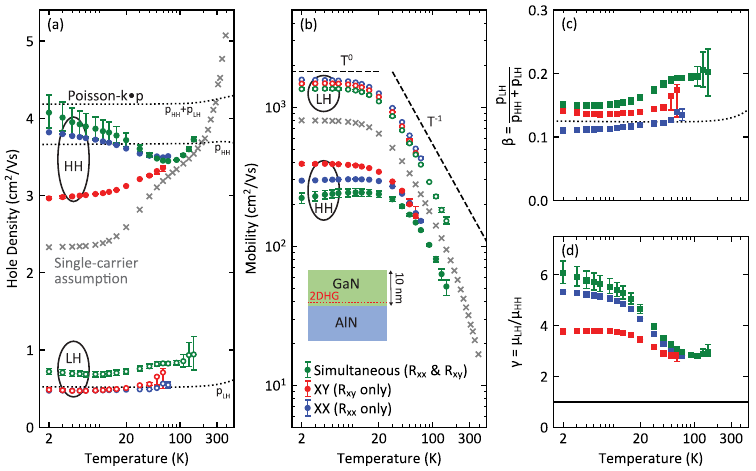}
    \caption{Measured LH (open circles) and HH (closed circles) (a) carrier density and (b) mobility, (c) $\beta=p_{LH}/(p_{HH}+p_{LH})$, and (d) $\gamma=\mu_{LH}/\mu_{HH}$ ratios from two-carrier analysis by least-squares fitting of $R_{xx}(B)$ (blue) and $R_{xy}(B)$ (red) measurements, and both simultaneously (green). Grey X's denote the total density and mobility inferred from a single-carrier fitting of the low-field measurements. Black dotted lines denote Poisson-k$\cdot$p simulation of the band-resolved densities. $T^0$ and $T^{-1}$ power laws are plotted in (b) for reference.}
    \label{fig: nmu_cj}
\end{figure*}

Fig. \ref{fig: nmu_cj} shows the GaN HH and LH ($q_1=q_2=+q)$ (a) densities ($p_{HH}=n_1$, $p_{LH}=n_2$) and (b) mobilities ($\mu_{HH}=\mu_1$, $\mu_{LH}=\mu_2$), and the corresponding values of (c) $\beta=p_{LH}/(p_{HH}+p_{LH})$ and (d) $\gamma=\mu_{LH}/\mu_{HH}$ obtained from the XX fit (blue), XY fit (red) and simultaneous fit (green). As in Fig. \ref{fig: mr_data_cj}c, error bars indicate the bounds of the parameter space in which $\chi^{RMS}$ is within 10\% of the minimum. Each fit is plotted until the temperature at which the lower bound of one of the free parameters lies below the noise line (\SI{350}{\milli\ohm}).

In Fig. \ref{fig: nmu_cj}a, the band-resolved densities are corroborated against a self-consistent Poisson and multi-band k$\cdot$p simulation of the heterostructure's band profile and valence band structure. The details of this model are discussed elsewhere \cite{Bader2019Wurtzite, Birner2011}. Input parameters were obtained from Ref. \citenum{Vurgaftman2003}, accounting for temperature dependence in the lattice\cite{Figge2009} and elastic\cite{Dai2016} constants of GaN and AlN. The simulated band-resolved carrier densities are plotted in \ref{fig: nmu_cj}a and \ref{fig: nmu_cj}c. 

The XX and XY fits both report $p_{LH}\sim$\SI[per-mode=reciprocal]{0.5e13}{\per\centi\meter\squared}, in agreement with the simulation. $p_{HH}$ from the XY fit increases from \SI[per-mode=reciprocal]{3.0e13}{\per\centi\meter\squared} at 2 K to \SI[per-mode=reciprocal]{3.4e13}{\per\centi\meter\squared} at 90 K, while $p_{HH}$ from the XX and simultaneous fits decreases to the same from $\sim$\SI[per-mode=reciprocal]{4e13}{\per\centi\meter\squared} at 2 K. The discrepancy in the results of these fitting schemes points to additional sources of magnetotransport not captured in this two-carrier model.

By contrast, the carrier density inferred from \text{single-carrier} analysis of the low-field resistances (open grey circles) decreases by almost 50\% between \SI{300}{\kelvin} and \SI{2}{\kelvin}, similar to the Hall results reported previously for various GaN 2DHGs\cite{Liu2013, Li2013, Chaudhuri2019, Zhang2021, Chaudhuri2022a, Chaudhuri2022, Beckmann2022}. This drop coincides with a roughly two-fold increase in $\gamma$ over the same temperature range. Inspecting Fig. \ref{fig: Hall_error} and Eq. \ref{eq: Rxy_expanded}, such an increase in $\gamma$ at fixed $\beta$ would produce an anomalous decrease in the apparent carrier density. Thus, we assert that the previously reported temperature-dependent carrier density below room temperature is fully explainable as an artifact of parallel conduction.

Above room temperature, however, the density inferred from single-carrier analysis increases to $\sim\SI{5e13}{\per\centi\meter\squared}$ near \SI{400}{\kelvin}, well above the simulated value. Ref. \citenum{Chaudhuri2022} reports a similar approximately linear increase in the carrier density of a GaN/AlN 2DHG with increasing temperature, reaching $\sim\SI{1e14}{\per\centi\meter\squared}$ near \SI{800}{\kelvin}, albeit using a single-carrier interpretation. Under the assumptions of the model derived in this report --- isotropic transport, drift mobility $\approx$ Hall mobility --- the single-carrier density is \textit{always} an underestimate of the total density (per Eq. \ref{eq: n_ap} and Fig. \ref{fig: Hall_error}). Thus, these measurements imply an \textit{actual} increase in the GaN 2DHG density above room temperature. As mentioned earlier, no definitive explanation yet exists as to a temperature-dependent source or sink of charges. However, understanding the origin of these carriers may prove crucial for the future utilization of GaN 2DHGs in high-temperature applications.

The LH and HH mobilities in Fig. \ref{fig: nmu_cj}b saturate at temperatures below $\SI{20}{\kelvin}$, indicating extrinsically-limited transport. Between the three fits, HH mobilities measured at 2 K range from 222-393 cm$^2$/Vs, and LH mobilities range from 1350-1580 cm$^2$/Vs. In Fig. \ref{fig: nmu_cj}d, the $\mu_{LH}/\mu_{HH}$ ratio saturates between 3.7 and 6.1 near 10 K. 

Above $\SI{20}{\kelvin}$, the mobilities are phonon-limited and decrease with a $\sim T^{-1}$ power law. This trend is consistent with simulations of hole mobility in GaN/AlN 2DHGs\cite{Bader2019Wurtzite} which show that room temperature transport is primarily limited by acoustic phonon scattering, which goes as $\mu_{2d}^{ADP}\sim\big((m^*)^2k_BT\big)^{-1}$ in two dimensions\cite{Davies1997} ($m^*$ is the effective mass). In the same temperature range, the mobility ratio (in Fig. \ref{fig: nmu_cj}d) approaches $\gamma=\mu_{LH}/\mu_{HH}\sim\SI{2.9}{}$. Although $\gamma$ is not discernable at room temperature due to the suppression of $j>2$ polynomial features, this saturation suggests that $\gamma$ will be greater than 1. Thus, even if $R_{xy}(B)$ appears linear within the measured magnetic field range, single-carrier estimates of the ensemble hole density and mobility are still subject to the errors plotted in Fig. \ref{fig: Hall_error}.

\subsection{Fitting Discrepancies and Error Analysis}\label{subsec: fitting and error}
\begin{figure*}
    \centering
    \includegraphics[width=\textwidth]{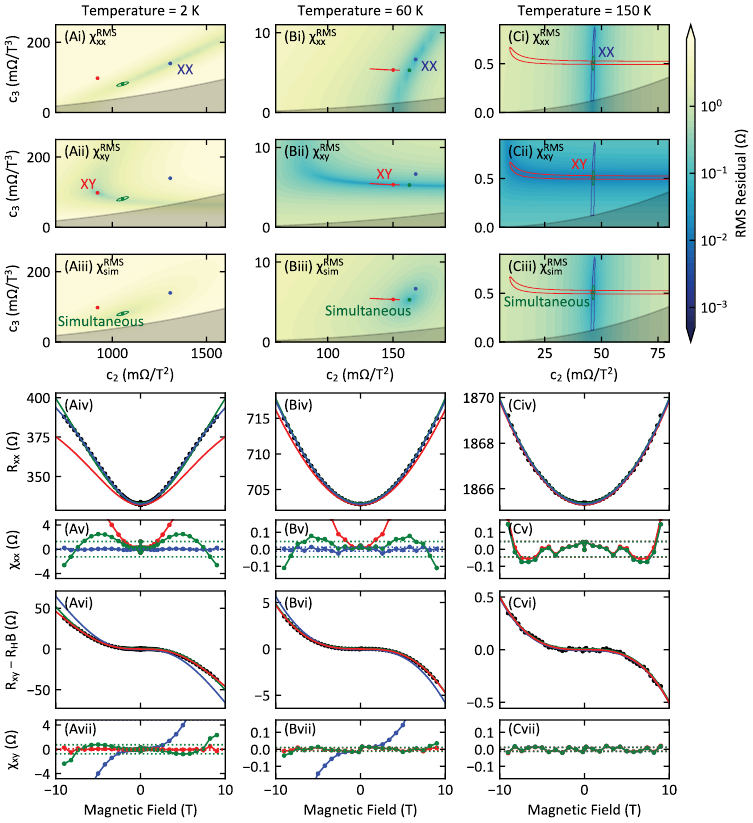}
    \caption{Demonstration of least-squares fitting at representative temperatures (A) 2 K, (B) 60 K, and (C) 150 K. Density plots in ($c_2$, $c_3$) space of the RMS residual of the two-carrier model against (i) $R_{xx}(B)$ and (ii) $R_{xy}(B)$ measurements, and (iii) both simultaneously. Contours enclose the regions in which $\chi^{RMS}$ is within 10\% of the minimum when fitting $R_{xx}(B)$ (blue), $R_{xy}(B)$ (red), and both simultaneously (green). (iv) Measured $R_{xx}(B)$ and (vi) $R_{xy}(B)-R_HB$ and the associated model results for the best-fit parameters in (i-iii), with the corresponding points-wise (connected points) and RMS (horizontal dotted line) residuals plotted in (v) and (vii), respectively.}
    \label{fig: resid_plots}
\end{figure*}
The nature of the discrepancy between the XX, XY, and simultaneous fit results and the size of the error bounds on the fit parameters are demonstrated in Fig. \ref{fig: resid_plots}. This discrepancy manifests in the differing determinations of $n_1$, $n_2$, $\mu_1$, and $\mu_2$ seen at low temperature in Fig. \ref{fig: nmu_cj}, and points to physics not contained in the assumptions of the model.

Model fitting at representative temperatures $T$ = 2 K, 60 K, and 150 K are plotted in columns A-C, respectively. The top rows of the figure show heat-density plots of $\chi^{RMS}$ (equations given in Table \ref{tab: fit}) for the (i) XX fit, (ii) XY fit, and (iii) simultaneous fit. $c_0$ and $c_1$ are equal in all plots at a given temperature, while $c_2$ and $c_3$ are varied to encompass the three best-fit regions. The plotted contours encompass all points in ($c_2$,$c_3$) where $\chi^{RMS}$ is within 10\% of the minimum. 

The darkened area in the lower-right corner of these plots denotes the region below $c_3=c_2^2/c_1$, where the band-resolved densities change sign (see the denominator in Eqs. \ref{eq: n1 from cj} and \ref{eq: n2 from cj}).

Row (iv) plots the measured $R_{xx}(B)$ and associated model result (Eq. \ref{eq: Rxx_cj}) for the best-fit parameters in the plots above. The point-wise (connected points) and RMS (dotted line) residuals for each fit are plotted versus magnetic field in row (v). Row (vi) similarly plots the measured and modeled $R_{xy}(B)-R_HB$, with the residuals plotted below in (vii).

The region of best-fit (lowest $\chi^{RMS}_{xx}$) for $R_{xx}$ (row i) appears as a diagonal line in ($c_2$,$c_3$) space at 2 K but becomes vertical at higher temperature as the $4^{\text{th}}$-order features are suppressed and cannot dictate the value of $c_3$ (by Eq. \ref{eq: cj}). Consequently, the contour of best fit in Fig. \ref{fig: nmu_cj}Ci becomes large, as do the error bars on $c_4$ in Fig. \ref{fig: mr_data_cj}c.

Similarly, the best-fit (lowest $\chi^{RMS}_{xy}$) region for $R_{xy}$ (row ii) draws an "L"-shaped contour in ($c_2$,$c_3$) space. At lower temperatures, the XY best-fit sits higher on the "L," indicating a strong $c_5B^5$ signal (again, see Eq. \ref{eq: cj}), while at higher temperatures, the XY best-fit sits in the horizontal region. 

At 2 K, the XX and XY best-fit regions differ by $\sim20\%$ in $c_2$ and $c_3$. This discrepancy is seen visually in the model fits (Aiv, Avi) and the trend in the residuals of the XY fit against the $R_{xx}$ data (in Av), and the residuals of the XX fit against the $R_{xy}$ data (in Avii). Consequently, the simultaneous fit is an imperfect representation of both the $R_{xx}(B)$ and $R_{xy}(B)$ measurements, and a trend is seen in the residuals of both datasets (see Av and Avi). 

In the supplementary material, Figs. 2-4 are recreated for a different GaN/AlN 2DHG sample capped with Mg-doped p-type GaN and measured in a Hall bar geometry. The same qualitative disagreement between the XX, XY, and simultaneous fits is observed in these measurements, which signals that the discrepancy between the XX and XY fits arises from an additional source of magnetoresistance independent of parallel conduction.


\section{Conclusions}
%
The large polarization-induced hole density at the GaN/AlN hetero-interface degenerately occupies both the HH and LH bands of the GaN; thus, Hall effect measurements must be interpreted using a proper multi-carrier model. 

This work describes an isotropic Drude model of two parallel-conducting carrier populations with the same charge polarity. We apply this model to fit 2DHG transport measurements from \SIrange{2}{170}{K} across a magnetic field range of $\pm$\SI{9}{\tesla}. We successfully resolve HH and LH carrier densities that are of comparable magnitude to self-consistent Poisson-k$\cdot$p simulations, as well as band-resolved mobilities, observing cryogenic mobilities of $\mu_{HH}\sim300$ cm$^2$/Vs and an LH mobility $\mu_{LH}\sim1400$ cm$^2$/Vs. Although it does not represent ensemble conduction, this LH mobility is the largest hole mobility reported to date in GaN.

Below \SI{20}{\kelvin}, the two-carrier model is unable to fit both the $R_{xx}(B)$ and $R_{xy}(B)$ measurements simultaneously without a trend in the residuals. An equivalent disagreement is seen in a similar GaN 2DHG structure in a different measurement geometry (see supplementary material), which suggests that this discrepancy arises from an additional source of magnetoresistance.

These results offer a credible experimental signature of LH band occupation in GaN and elucidate previously reported temperature-dependent transport results in GaN 2DHGs, which suggested an unexpected decrease in carrier density with decreasing temperature. In future transport studies of GaN 2DHGs (or similar systems with two or more parallel-conducting channels), if the sample's sheet resistance is too large or sufficiently large magnetic fields are not available to observe higher-order polynomial features in $R_{xx}(B)$ and $R_{xy}(B)$, transport simulations should predict and be corroborated against measurements of $R_{sh}$ and $R_H$, which are directly measurable, rather than $n$ and $\mu$.

\section{Supplementary material}
In the supplementary material, Figs. \ref{fig: mr_data_cj}, \ref{fig: nmu_cj}, and \ref{fig: resid_plots} are recreated for magnetotransport measurements of a GaN/AlN 2DHG sample with a Mg-doped p-type GaN cap, measured in a Hall bar geometry up to $\pm$9 T from 3 K to 390 K.

\section{Acknowledgements}
This work was supported in part by SUPREME, one of seven centers in JUMP 2.0, a Semiconductor Research Corporation (SRC) program sponsored by DARPA, and as part of the Ultra Materials for a
Resilient Energy Grid (epitaxy and device fabrication), an Energy
Frontier Research Center funded by the US Department of Energy,
Office of Science, Basic Energy Sciences under Award No. DESC0021230 and in part by the ARO Grant No. W911NF2220177
(characterization). The authors acknowledge the use of facilities and instrumentation supported by NSF through the Cornell University Materials Research Science and Engineering Center DMR-1719875, and the use of the Cornell NanoScale Facility (CNF), a member of the National Nanotechnology Coordinated Infrastructure (NNCI), which was supported by the National Science Foundation (NSF Grant No. NNCI-2025233). The authors benefitted from discussions with Jie-Cheng Chen, Feliciano Giustino, Changkai Yu, Samuel James Bader, and Reet Chaudhuri.

\section{Author Declarations}
\subsection{Conflict of Interest}
The authors have no conflicts to disclose.

\subsection{Author Contributions}
\textbf{Joseph E. Dill}: Investigation (equal), methodology (lead), visualization (lead), writing - original draft (lead); \textbf{Chuan F. C. Chang}: Conceptualization (lead), data curation (lead), investigation (equal), writing - review \& editing (equal); \textbf{Debdeep Jena}: Supervision (supporting), writing - review \& editing (equal), Funding Acquisition (equal). \textbf{Huili (Grace) Xing}: Supervision (lead), writing - review \& editing (equal), Funding Acquisition (equal).

\section{Data Availability Statement}
The data that support the findings of this study are available from the corresponding author upon reasonable request. 

\section{References}
\bibliography{references}

\end{document}


\preprint{Supplementary Material}

\title{
Supplementary Material \\Two-Carrier Model-Fitting of Hall Effect in Semiconductors with Dual-Band Occupation: A Case Study in GaN Two-Dimensional Hole Gas
}
\author{Joseph E. Dill*}
 \affiliation{School of Applied \& Engineering Physics, Cornell University, Ithaca, New York 14853, USA}
\author{Chuan F. C. Chang*}%
\affiliation{The Department of Physics, Cornell University, Ithaca, New York 14853, USA}%
\author{Debdeep Jena}
\author{Huili Grace Xing}
\affiliation{Electrical and Computer Engineering, Cornell University, Ithaca, New York 14853, USA}
\affiliation{Materials Science and Engineering, Cornell University, Ithaca, New York 14853, USA}
\affiliation{Kavli Institute at Cornell for Nanoscale Science, Cornell University, Ithaca, New York 14853, USA\\ *These two authors contributed equally to this work.}
\date{\today}

\maketitle

\begin{figure*}
    \centering
    \includegraphics[width=\textwidth]{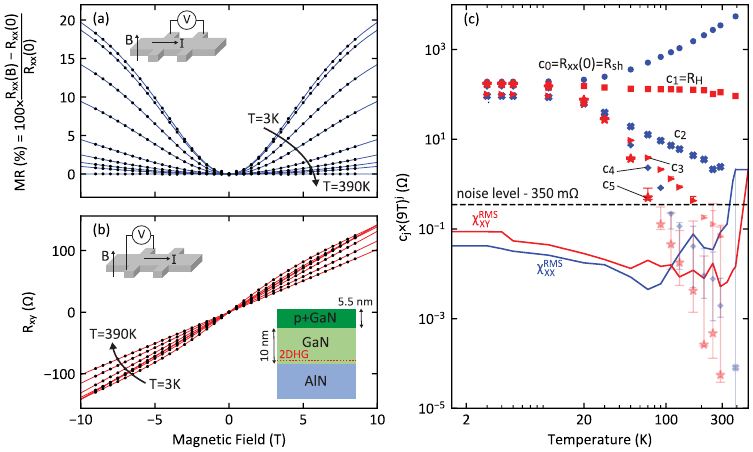}
    \caption{Measured (a) MR$(B)$ and (b) $R_{xy}(B)$ of a GaN/AlN 2DHG, field-swept to $B=\pm$\SI{9}{T} at select temperatures from \SIrange{2}{390}{K}. The four-point measurement configuration is shown for each. (c) The RMS residuals of the fits in (a) $\chi_{xx}^{RMS}$ (blue line) and (b) $\chi_{xy}^{RMS}$ (red line) versus temperature, and the signal contributions of the $j$-th polynomial features of $R_{xx}(B)$ (even $n$) or $R_{xy}(B)$ (odd $n$) at $B$=9 T (the maximum measured field) for $n=0$ (circles), $n=1$ (squares), $n=2$ (X's), $n=3$ (triangles), $n=4$ (diamonds), and $n=5$ (stars). Points with an uncertainty bound below the \SI{350}{\milli\ohm} noise line are dimmed.
    }
    \label{sup fig: mr_data_cj}
\end{figure*}
\begin{figure*}
    \centering
    \includegraphics[width=\textwidth]{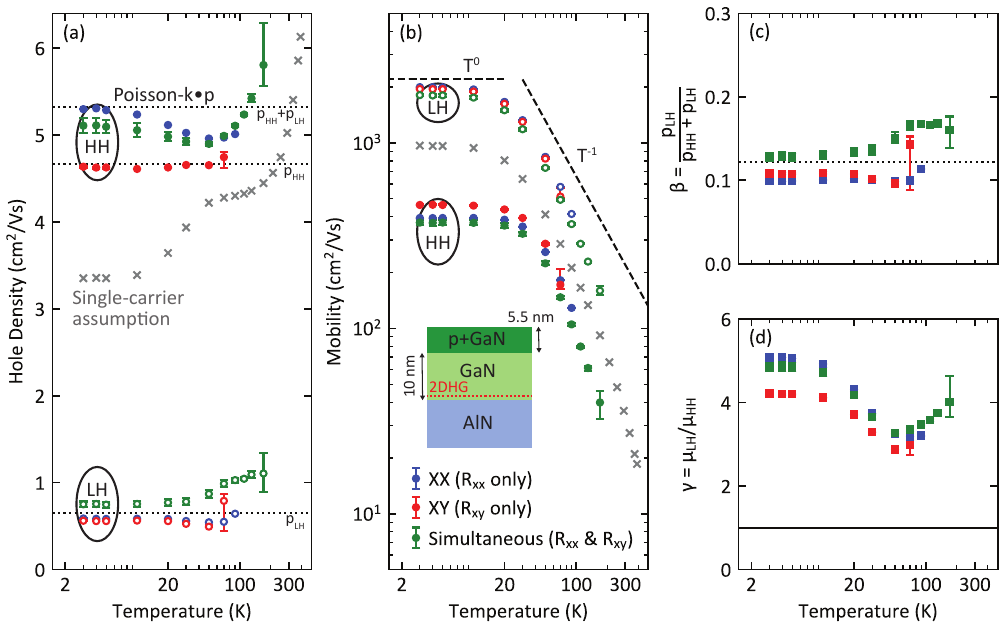}
    \caption{Measured LH and HH (a) mobility and (b) carrier density, (c) $\beta=p_{LH}/(p_{HH}+p_{LH})$, and (d) $\gamma=\mu_{LH}/\mu_{HH}$ ratios from two-carrier analysis by least-squares fitting of $R_{xx}(B)$ (blue) and $R_{xy}(B)$ (red) measurements, and both simultaneously (green). Self-consistent Poisson-k$\cdot$p simulations (black dotted line) are also plotted.}
    \label{sup fig: nmu_cj}
\end{figure*}
\begin{figure*}
    \centering
    \includegraphics[width=\textwidth]{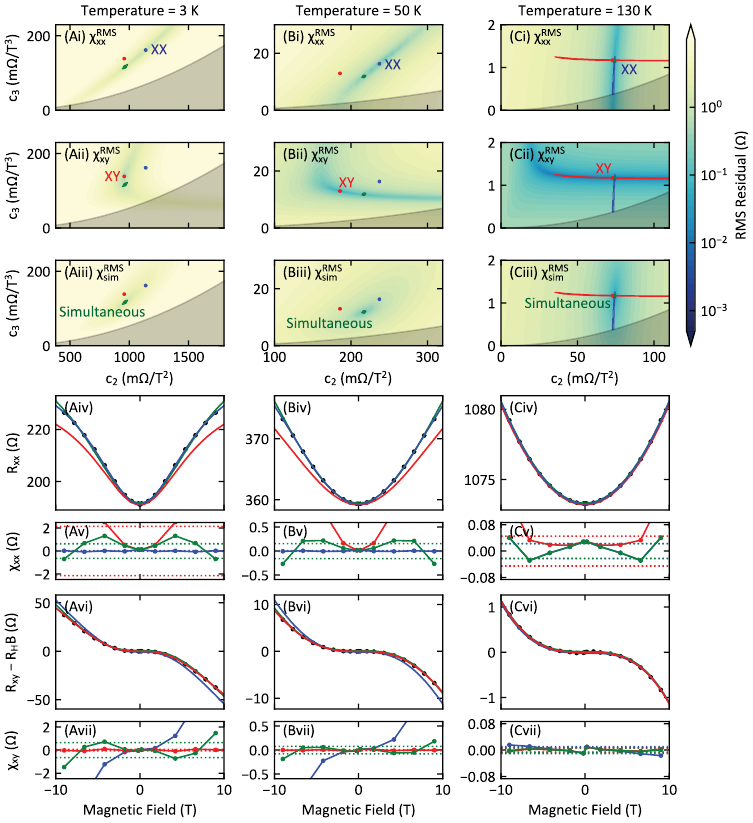}
    \caption{Least-squares fitting at representative temperatures (A) 3 K, (B) 50 K, and (C) 130 K. Density plots in ($c_2$, $c_3$) space of the RMS residual of the two-carrier model against (i) $R_{xx}(B)$ and (ii) $R_{xy}(B)$ measurements, and (iii) both simultaneously. Contours enclose the regions in which $\chi^{RMS}$ is within 10\% of the minimum when fitting $R_{xx}(B)$ (blue), $R_{xy}(B)$ (red), and both simultaneously (green). (iv) Measured $R_{xx}(B)$ and (vi) $R_{xy}(B)-R_HB$ and the associated model results for the best-fit parameters in (i-iii), with the corresponding points-wise (connected points) and RMS (horizontal dotted line) residuals plotted in (v) and (vii), respectively.}
    \label{sup fig: resid_plots}
\end{figure*}
\begin{widetext}

In this supplementary material, Figs. 2-4 from the main text are recreated for a second GaN/AlN 2DHG sample. This sample, also grown by molecular beam epitaxy, consisted of a \SI{5.5}{\nano\meter} layer of p-type GaN doped with $\sim$\SI{1e19}{\per\centi\meter\cubed} magnesium, then a \SI{10}{\nano\meter} layer of undoped GaN with on \SI{700}{\nano\meter} of undoped AlN, grown on a single-crystal aluminum nitride substrate (see Fig. \ref{sup fig: mr_data_cj}b). The magnetotransport measurements were performed in the same way as described in the main text, except that a lithographically-defined hall bar geometry was used (electrical configurations shown in insets to Fig. \ref{sup fig: mr_data_cj}a and \ref{sup fig: mr_data_cj}b).

The contents of Figs. \ref{sup fig: mr_data_cj}, \ref{sup fig: nmu_cj}, and \ref{sup fig: resid_plots} are equivalent to those of Figs. 2, 3, and 4 of the main text, respectively. In Fig. \ref{sup fig: resid_plots}, the nature of deviation between the XX, XY, and simultaneous fits is equivalent to that seen in Fig. 4 of the main text, namely the $R_{xx}(B)$ data are best fit with a larger $c_2$ and $c_3$ than the $R_{xy}(B)$ data, resulting in a larger estimate of the HH density and a lower estimate of the HH mobility at all temperature below 100 K (see Figs. \ref{sup fig: nmu_cj}). The fact that this deviation persists across samples and measurement geometries implies that it is not an artifact of the measurement but arises from a source of magnetoresistance besides parallel conduction.

Another point of similarity between the two samples is the apparent carrier density from single-carrier analysis (grey X's in Fig. \ref{sup fig: nmu_cj}b), which rises to \SI[per-mode=reciprocal]{6e13}{\per\centi\meter\squared} at \SI{400}{\kelvin}, increasing sharply above the Poisson-k$\cdot$p simulation of the carrier density at the same temperature. As discussed in the main text, single-carrier analysis always \textit{under}-estimates carrier density, so these measurements imply an \textit{actual} increase in the GaN 2DHG density above room temperature, which is not yet understood.
\end{widetext}